\documentstyle[prb,twocolumn,aps]{revtex}
\newcommand{\beq}{\begin{equation}}
\newcommand{\eeq}{\end{equation}}

\def\half{{1\over2}}

\def\rhob{{\bar \rho}}

\def\eqa{\begin{eqnarray}}
\def\eea{\end{eqnarray}}
\begin{document}
\input psfig
\draft
\flushbottom
\twocolumn[
\hsize\textwidth\columnwidth\hsize\csname @twocolumnfalse\endcsname
\title{  Scaling Relations for  Gaps in  Fractional Quantum
Hall States}
\author{  Ganpathy Murthy$^{a}$, K.Park$^{b}$, R.Shankar$^{c}$, and
J.K.Jain$^{b}$  }
\address{
{\it (a)} Physics Department, Boston University, Boston MA 02215\\
and
Department of Physics and Astronomy, Johns Hopkins University,
Baltimore MD 21218;\\
{\it (b)} Department of Physics and Astronomy, State University of New York  
at Stony Brook,
Stony Brook, NY 11794\\
{\it (c)} Department of Physics, Yale
University, New Haven CT 06520}
\date{\today}
\maketitle
\tightenlines
\widetext
\advance\leftskip by 57pt
\advance\rightskip by 57pt

\begin{abstract}
The microscopic Hamiltonian approach of Murthy and Shankar, which has
recently been used to calculate  the transport gaps of quantum Hall
states with fractions $\nu = {p \over 2ps +1}$, also implies
scaling relations between gaps within a single sequence (fixed $s$)
as well as between gaps of corresponding states in different sequences.  
This work tests these  relations for a system of electrons in the
lowest Landau level interacting with a model
potential cutoff at high momenta due to sample thickness.
\end{abstract}
%\vskip 1cm
\pacs{71.10.Pm, 73.40.Hm, 73.50.Jt, 74.20.-z}
]

\narrowtext
\tightenlines
In the theory of the Fractional Quantum Hall\cite{fqhe-ex} states,
composite fermions (CFs)\cite{jain-cf} play a central role. It was
pointed out by Jain that for fractions $\nu = {p \over 2ps +1}$ if  
one
traded the electrons for CFs carrying $2s$ units of statistical flux,
the CF would  see a weaker field $B^* = B/(2ps +1)$, which would
then be just right to fill exactly $p$ CF-Landau levels (CF-LLs).  
Starting
with $\Phi_p$,  the fermionic wavefunction for the $p$ filled LL  
state,
Jain deduced the electronic wavefunction at $\nu = {p \over 2ps +1}$
by multiplying by Jastrow factor $\Phi_1^{2s}$, where $\Phi_1$ is the
wave function at $\nu=1$, and projecting to the
lowest Landau level (LLL) to obtain ${\cal P} \Phi_1^{2s} \Phi_p$,    
the FQHE
wave function in terms of electronic coordinates. This procedure has  
yielded
excellent
wavefunctions for the ground and low lying excited
states\cite{jain-cf,jain-cf-review} and gaps
 shown to be accurate to within a   few percent  
level\cite{jain-cf-review}.

Turning from  successful trial wavefunctions to the quest for   a   
hamiltonian or functional description that links the original 
electronic problem to  the final low energy
physics through some type of approximation,  we encounter the  
Chern-Simons (CS) 
theory\cite{CS} which allows one to trade
electrons for bosons\cite{bosons} or fermions\cite{lopez}
carrying varying amounts of flux for both gapped\cite{lopez} and
gapless\cite{kalmeyer,hlr} states.  A hamiltonian description
within this approach  put forth by Murthy and  
Shankar (MS)\cite{us1,us2},
 was recently used to calculate the gaps of various fractional
quantum Hall states\cite{single-part}.
The objective of the present paper is to test not just the magnitudes  
of the
excitation gaps for various fractions  but  the relationships between  
them
implied by the
MS approach.

In the MS approach  the CS flux attachment is
followed by an enlargement of the Hilbert space to include $n$
magnetoplasmon oscillator degrees of freedom, $n$ being the number of
electrons per unit area, following an old idea of Bohm and
Pines\cite{bohm-pines}. There arise $n$ constraints which  preserve  
the
degrees
of freedom.  Upon
decoupling the oscillators from the fermions in the {\em infrared}  
one obtains
a description of CFs  which has many
desired attributes.  For example, the effective $1/m^*$ vanishes in
the noninteracting limit and owes its existence to the interactions,
 and an  effective magnetic moment that couples to
 an inhomogeneous magnetic field\cite{ssh} emerges very
naturally\cite{us2}. Most importantly,  the electronic charge in  
terms of the
final CF coordinates and momenta takes the form:

\eqa
\rho^e(q)&=&{q \over \sqrt{8\pi}} \sqrt{{2p\over2ps+1}} (A(q)+
A^{\dagger} (-q)) \nonumber \\ &+&{\sum_j e^{-iqx_j} \over 2ps+1}
-{il^{2} } \sum_j (q \times \Pi_j)e^{-iqx_j}\label{rhobar}
\eea where
 $l=1/\sqrt{eB}$ is the
magnetic length, and ${\vec\Pi}_j={\vec P}_j+e{\vec A}^*(r_j)$ is the
velocity operator in the effective field.  The oscillator piece  
(first
line) saturates Kohn's theorem\cite{kohn}, and the remaining
low-energy piece (the last two terms),
which we henceforth call $\rhob$, satisfies the magnetic translation
algebra\cite{GMP} to lowest leading order. Note that $\rhob$ is a sum
of a monopole with the correct quasiparticle\cite{laughlin,jain-cf}  
charge
$e^*=e/(2ps+1)$, and a dipole
which alone survives at $\nu=\half$ and has the value obtained
previously \cite{read2} (see also
Refs.~\onlinecite{dh,pasquier,read3}). Finally, the ratio of the monopole to  
dipole
pieces is such as to give transition matrix elements of order $q^2$.   
Note that
we trust our
expressions only for small $ql$.

Given $\bar{\rho}$, one can construct the  low-energy CF hamiltonian
(suppressing  the  magnetic moment term not germane here)
\beq H=\half \int {d^2
q\over(2\pi)^2} v(q) \rhob(-q) \rhob(q). \label{ham}
\eeq
The $n$ constraints act on just the fermions when the oscillators are
decoupled.
Given that many
nonperturbative affects (like mass and charge renormalization) are
built in, we can expect approximation methods to be quite effective.

If one knew the exact eigenstates of $H$ one could infer  the gap  
from  the
difference in the expectation value $\langle H \rangle$  in the  
ground state
and a state with a widely separated particle-hole pair. We now turn  
to  some
obstacles that arise in practice  and their treatment.

1. {\em The exact eigenstates of $H$ are not known.}  In their place
approximate ones that
 describe the noninteracting (single-particle) part of $H$ are  
employed.  The
state with $p$ filled CF-LLs and
particle-hole excitations of it will serve as the ground state and  
the
particle-hole states in which  $H$ is
averaged. Since these happen to be   Hartree-Fock eigenstates of
$H$, they are  a good starting
point for perturbative approximations\cite{single-part}.

Note that
these wavefunctions are not multiplied by Jastrow factors
and projected  to the LLL as in Jain's work: Instead one uses  
$\bar{\rho}$ of
Eqn.(\ref{rhobar}) for the  projected electronic density in the final  
CF
coordinates, which contains the same
physics.  This difference in the representation  describing composite
fermions in the Jain  and the MS approaches must be
emphasized.  In the former,  the Hamiltonian and density  
operators
remain unchanged in form, the electronic density for example  being   
given
$\rho (q) = \sum_j e^{-iqx_j}$ where $x_j$ is  the electronic  
coordinate.  In
the
MS approach, the effect of vortex attachment and projection are
incorporated by the  sequence of transformations of the
electron density, and the  freezing of oscillator degrees of freedom.  
Whether
one  uses the simple operators and complicated wavefunctions (Jain)  
or vice
versa (MS)  amounts to a choice of representation.

2.  {\em The expression for $\bar{\rho}$ and hence $H$ is to be
trusted only for small $ql$. } This turns out to be only a minor
handicap for realistic high-density samples. We approximate the
effects of sample thickness by using the following model
potential\cite{thick2}
\beq
v(r)={e^2\over \sqrt{r^2+\Lambda^2}},
 \ \ \ \ \ \ \ \  v(q)={2\pi e^2\over q} e^{-\Lambda q}\;,
 \eeq
where $\Lambda$ is a parameter related to the thickness. For large
$\Lambda/l$, matrix elements of $\bar{\rho}$ are needed only at small  
$q$.

3.  {\em There are $n$ constraints on the physical states. } If one
sandwiches the exact gauge invariant charge operator between gauge
invariant eigenstates of the exact hamiltonian, constraints will not
matter, since the restriction to physical intermediate states will be
automatic. In the present case, $\bar{\rho}$ is expected to be gauge
invariant, since it obeys the small $q$ magnetic algebra\cite{GMP}
 when the commutators are evaluated in the full space
without regard to constraints. However, there is the problem that the
Hartree-Fock CF states used are not gauge invariant.  The approximate
way the constraints were handled is described in
Ref.~\onlinecite{single-part}. One choice, followed here, is to ignore them.   
Although
they are not explicitly taken into account, they were invoked in  
deriving
Eqn.(\ref{rhobar}), as explained in Ref.~\onlinecite{us1}. (Constraints  
can be
very important at and near the gapless fraction  $\nu =1/2$. For a  
discussion
see Refs.~\onlinecite{comment,read3}.)

Since we are going to move between various fractions $\nu =  
p/(2ps+1)$ it is worth
deciding what varies and what remains fixed.  We will always consider  
systems
with fixed density $n$, varying the field to change the filling. The magnetic  
lengths
$l$  and $l^*$ of the electrons and CF respectively are then given by
\begin{equation}
l = { 1 \over \sqrt{eB}} = \sqrt{{p\over  2\pi n(2ps+1)}} \ \ \ \ \  
\
l^* = { 1 \over \sqrt{eB^*}} = \sqrt{{p\over  2\pi n}}
\end{equation}
{\em The fact that  $l^*$ is independent of $s$ will play a crucial  
role.}

Within the MS
scheme, $\Delta(\Lambda,{p\over 2ps +1})$, the gap at 
$\Lambda$ and $\nu = p/(2ps +1)$ is
given by an expression
\begin{eqnarray}
\Delta\left({p\over 2ps +1},\Lambda\right) &=&  \int {d^2 q\over  
(2\pi  )^2} {
\pi e^2
\over q}
e^{-q\Lambda} \nonumber \\
& & \!\!\!\!\!\!\!  \!\!\!\!\!\!\! \!\!\!\!\!\!\!  \!\!\!\!\!\!\!
\times \sum_n
 \left[  |\langle n|\bar{\rho}(q)|  \Phi_p + PH\rangle|^2 -
|\langle
n|\bar{\rho}(q)| \Phi_p\rangle|^2\right] \label{gapexp}
\end{eqnarray}
where $\Phi_p$ is the free particle ground state with $p$ filled  
CF-LLs,
$\Phi_p + PH$ is the ground state plus a widely separated  
particle-hole
pair, and $\sum_n$ is the sum over all CF states. The gaps were  
computed in
Ref.~\onlinecite{single-part} using this formula.

In addition to yielding  absolute numbers, Eqn. (\ref{gapexp})   
yields
relations between gaps for
different fractions. To this end, let us turn to
 $\bar{\rho}$ in Eqn.(\ref{rhobar}), use  $l^2= l^{*2}/(2ps+1)$, and    
write
\begin{eqnarray}
\bar{\rho} &=&{1 \over 2ps+1}\left[ \sum_j e^{-iqx_j}  -{il^{*2} }   
\sum_j
(q \times
\Pi_j)e^{-iqx_j}\right] \\
& \equiv & {1 \over 2ps+1} \rho^* (q).\label{rhostar}
\end{eqnarray}
{\em The entire $s$ dependence of the gap is in the explicit factor  
of
$1/(2ps+1)^2$  relating $\bar{\rho}^2 $ to $\rho^{*2}$. Both     
$\rho^*$, and
its matrix elements in the states referred to in  Eqn. (\ref{gapexp})  
depend
only  on $l^*$ and hence $p$, but not $s$.} Let us make this explicit  
by writing Eqn.(\ref{gapexp}) as
\begin{equation}
\Delta\left({p\over 2ps +1},\Lambda\right) = {1 \over (2ps+1)^2}    
\int {d^2
q\over (2\pi  )^2} { \pi e^2
\over q}
e^{-q\Lambda} f(ql^*)\label{basic}
\end{equation}

{\em Scaling Relation I}: Consider the limit $\Lambda \to \infty$. We  
can
replace $f(ql^*)$ by $f(0)$, which, according to 
Ref.~\onlinecite{single-part},  has
a value $2$ for any $p$,  to obtain
\begin{equation}
\Delta\left({p\over 2ps+1},\Lambda\right) \to
\frac{e^2}{\Lambda}\frac{1}{(2ps+1)^2}\;.
\label{gaplam}
\end{equation}
from which follows:
\begin{equation}
\lim_{ \Lambda \to \infty} {\Delta ({p_1\over 2p_1 s_1+1}, \Lambda_1  
)\over
\Delta ({p_2\over 2p_2 s_2+1}, \Lambda_2)} =
\left( {2p_2 s_2 +1\over 2p_1 s_1 +1}\right)^2 {\Lambda_2 \over
\Lambda_1}.\label{scaling1}
\end{equation}

So far the gap $\Delta$ and the thickness parameter 
$\Lambda$ have been defined in  some laboratory units, say, eV and 
Angstroms. We define a dimensionless thickness parameter 
\beq
{\lambda} = {\Lambda \over l}\;.
\eeq
It is a common practice to quote gaps in units of 
$e^2/l$, but here, since the gap behaves as $e^2/l$ for small
$\lambda$ and as $e^2/\Lambda$ for large $\lambda$, it proves more 
convenient to consider the dimensionless gap $\delta$ defined by
\beq
\Delta = {e^2 \over \sqrt{l^2 + \Lambda^2}}\; 
\delta({p \over 2ps+1}, \lambda )\;,
\label{delta}
\end{equation}
and viewed as a function of $\lambda$.
Then, for a given $\lambda$
\begin{equation}
\lim_{{\lambda}\to \infty} \;
{{\delta}({p_1\over 2p_1s_1 +1}, {\lambda} )\over
{\delta}({p_2\over 2p_2s_2 +1}, {\lambda})}=
\left( \frac{2p_2s_2 +1}{2p_1 s_1+1}\right)^2 .\label{scaling11}
\end{equation}
This will be called {\em Scaling Relation I.}

{\em Scaling Relation II}.  Let us return to Eqn. (\ref{basic}) and   
note that
since $f(ql^*)$ does not depend on $s$,
\begin{equation}
\Delta\bigg( {p\over 2ps_1 +1} , \Lambda \bigg)  =
\left[ {2ps_2+1 \over 2ps_1 +1
}\right]^{2}
\Delta\bigg({p\over 2ps_2 +1}  , \Lambda \bigg) .\label{scaling2}
\eeq
This is {\em Scaling Relation II.}  This too may be transcribed in
terms of $\delta$ using Eqn.~(\ref{delta}).

{\em At the heart of the scaling
relations are two simple facts: (i) The two parts of $\bar{\rho}$ --
monopole and dipole -- are such that we can scale out a factor
$1/(2ps+1)$ from both when we express $\bar{\rho}$ in terms of the
natural variable $ql^{*}$ and, (ii) The states used do not vary
with $s$. }  The careful reader will note that the same scaling
relation will follow if one had just the monopole piece of charge
$1/(2ps+1)$. This will, however, be unacceptable for other reasons:
The magnetic algebra\cite{GMP} requires the dipole piece, and equally
important, without the dipole piece, $\bar{\rho}$ will have  
transition
matrix elements linear in $q$, in violation of Kohn's
theorem\cite{kohn}. Indeed, with the present ratio of terms, the
linear terms in $q$ from the monopole and dipole pieces precisely
cancel.

We now test the preceding results, starting with the absolute values  
of the gaps. 
The expressions for the gaps in the MS approach have been given
earlier\cite{single-part}, and compared with the gaps obtained by  
Park and Jain\cite {park} from CF wave
functions.  In Fig.~1, we show a comparison for a larger range of
$\lambda$ for $\nu=1/3$, 2/5, and 3/7.  The reader is referred to the
literature for the details of the variational Monte Carlo methods  
used
for evaluating the gaps from the CF wave  
functions\cite{kamilla,park}.
The MS gap does not work at small $\lambda$, as expected, since  
here
the energetics is controlled by the large $q$ terms in the
Hamiltonian, which have been neglected in the above MS analysis.   It
becomes better until $\lambda\approx 3$, but worsens for larger
$\lambda$, possibly because  the spacing
between the first unoccupied and higher CF-LLs becomes small, making
fluctuation corrections to Hartree-Fock important.

Now we ask if  the scaling  relations between various gaps
implied by the MS theory may be more robust than the numerical
values of the gaps themselves.

We plot in Fig.~2 the ratios of gaps involving filling
fractions 1/3, 2/5, and 3/7, i.e.,  test  Scaling Relation I,
Eqn.(\ref{scaling11}). Two features are noteworthy. First, when we go  
beyond
${\lambda} >4 $, the ratios are fairly close to the asymptotic  
values of
$25/9 = 2.78 $ and $49/25 = 1.96$ computed in Hartree Fock.
Next, even at very small  thickness, the {\em ratios}  of the MS gaps  
agree well with the Monte Carlo results. The reason why the ratios of  
gaps come out much better than their absolute values at small $\lambda$  
is not understood at this point.

\begin{figure}
\centerline{\psfig{figure=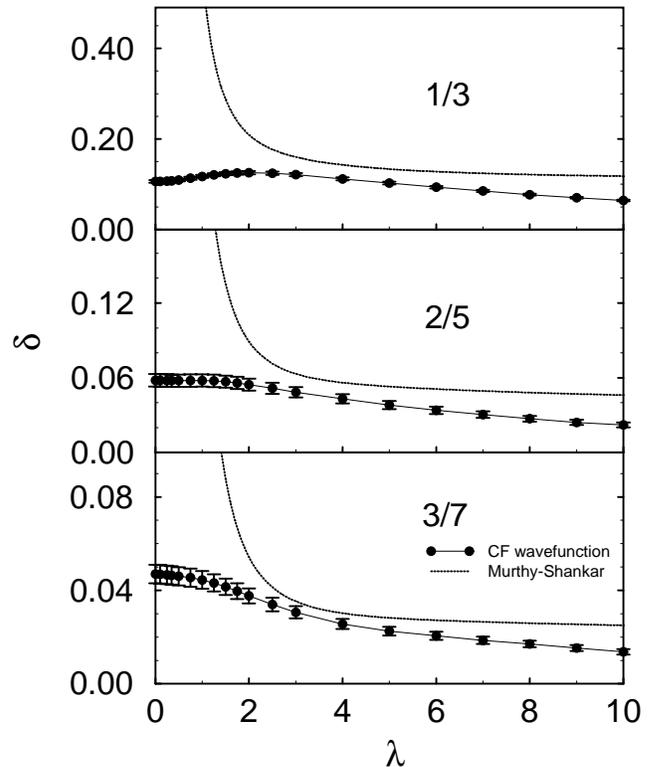,width=3.5in,angle=0}}
\caption{The excitation gaps (units defined in text)
for filling fractions 1/3, 2/5 and 3/7
computed from the CF wave functions for the model interaction potential
$v(r)$.  The error bars show the statistical uncertainty in Monte
Carlo.  The dashed lines are the gaps predicted by the MS approach.
\label{fig1}}
\end{figure}

\begin{figure}
\centerline{\psfig{figure=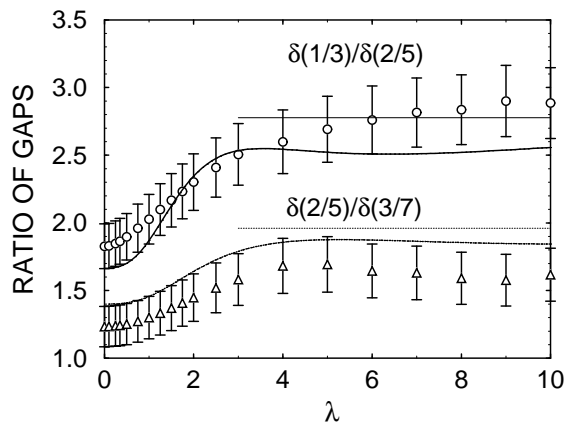,width=3.5in,angle=-90}}
\caption{Ratios of excitation gaps at filling fractions
1/3, 2/5 and 3/7.  The points are the calculations from the CF wave
functions and the lines from the MS formulas.  The horizontal lines
depict the asymptotic $\lambda \rightarrow \infty$ limit of the MS ratio.
\label{fig2}}
\end{figure}

We note that earlier work\cite{hlr,kamilla}
found that the gaps at $\lambda=0$ scale approximately as
\begin{equation}
\Delta=C \frac{e^2}{l} \frac{1}{(2p+1)}\;,
\end{equation}
which  implies that
\begin{equation}
\Delta\propto [p(2p+1)]^{-1/2}\;.
\end{equation}
A comparison with Eq.~(\ref{gaplam}) shows that 
the gaps at large $\lambda$ should decrease faster with $p$ 
than the ones at $\lambda=0$.  Another interesting implication
is regarding the effective mass of composite fermions, $m^*$, which  
is
obtained by equating the gaps to the CF cyclotron energy
$eB^*/m^*=eB/(2p+1)m^*$.  While at $\lambda=0$ one would expect
$m^*\sim \sqrt{(2p+1)/p} \sim \sqrt{B}$, at large $\lambda$,
Eq.~(\ref{gaplam}) implies $m^*\sim (2p+1)^2/p$, which has a much
stronger dependence on the filling factor, diverging as $\nu
\rightarrow 1/2$.  These results are generally consistent with
the trends found in the gaps computed from the CF wave
functions\cite{park}.

Now we turn to  Scaling Relation II, Eqn.(\ref{scaling2}).
Fig.~3 shows the scaling prediction for the gaps   
at $p/(4p+1)$, [which requires the gaps at $p/(2p+1)$ as input], 
along with the gaps  at
$p/(4p+1)$ computed directly from the CF wave functions. For large
values of $\lambda$, the two gaps are in agreement within numerical
uncertainty. But  they are very close  even for small $\lambda$,
which is unexpected in the formalism aimed at  the  infrared.

\begin{figure}
\centerline{\psfig{figure=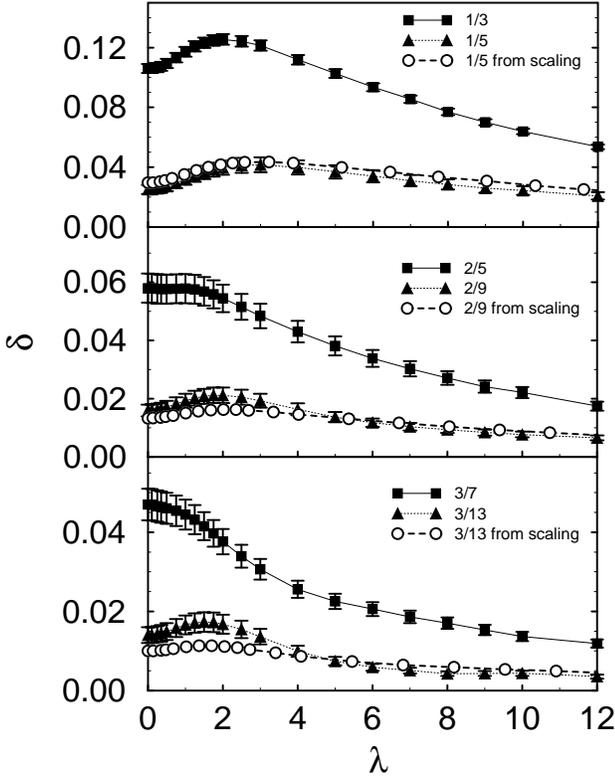,width=3.5in,angle=0}}
\caption{The gaps for several FQHE states at $p/(2p+1)$ and $p/(4p+1)$
computed from the CF wave functions for the model interaction
potential
$v(r)$.  The circles show gaps at
$p/(4p+1)$ deduced from the gaps at $p/(2p+1)$ using the scaling
relation discussed in the text.
\label{fig3}}
\end{figure}

In hindsight, one can motivate the above scaling
of gaps at corresponding fractions as follows.  It has been noted
before\cite{kamilla2} that the charge density profiles of the
particle or hole excitations at $p/(2sp+1)$ are largely independent  
of $s$, when plotted in terms of $l^*$. The  main difference is
that the total integrated charge is $e^*=e/(2sp+1)$.  Therefore, it  
is
natural to expect that the
excitation energy is proportional to $e^{*2}/l^*$.  (The constant of
proportionality will clearly depend on $p$, since the density  
profiles of
of the  excitations depend strongly on $p$, but may be expected to
have only a weak dependence on $s$). This is precisely the above
scaling relation.  While the argument is very natural, it not trivial
to derive it systematically within the theory.

We have neglected the possibility that
the FQHE  become unstable at large thicknesses,
as found in numerical studies\cite{Haldane}.
This physics is not relevant to our main concern here, which
is to test the consistency of the MS approach.

In conclusion, we have tested functional relations between the gaps
coming out of the
Hamiltonian approach of Murthy and Shankar, and found them to
be in good agreement with the results obtained from the CF wave
functions.  G.M.  thanks Andy Millis and R.S thanks N. Read for
stimulating conversations.  We are grateful for support from the NSF
grants no. DMR98-00626 (RS) and DMR-9615005 (JKJ) and
a grant of computing time on the SGI Power Challenge cluster at the  
NCSA,
University of Illinois, Urbana-Champaign.

\end{document}